\documentclass[journal]{IEEEtran}

\usepackage{ifpdf,cite}
\ifCLASSINFOpdf
   \usepackage[pdftex]{graphicx}
   \graphicspath{{../pdf/}{../jpeg/}}
   \DeclareGraphicsExtensions{.pdf,.jpeg,.png}
\else
   \usepackage[dvips]{graphicx}
   \graphicspath{{../eps/}}
   \DeclareGraphicsExtensions{.eps}
\fi
\usepackage{epsf,psfrag,epsfig}

\usepackage[cmex10]{amsmath}
\usepackage{mathrsfs, amsthm,amsfonts, latexsym, amssymb}
\usepackage[mathscr]{eucal}
\usepackage{graphicx}
\usepackage[breaklinks]{hyperref}
\usepackage[hyphenbreaks]{breakurl}
\usepackage[caption=false,font=scriptsize]{subfig}
\usepackage{color}
\usepackage{enumitem}
\theoremstyle{definition}
\newtheorem{mydef}{Definition}
\theoremstyle{plain}
\newtheorem{lem}{Lemma}
\newtheorem{thm}{Theorem}
\IEEEoverridecommandlockouts
\begin{document}
\title{Probabilistic Forecasting of Real-Time LMP and Network Congestion}

\author{Yuting~Ji,~\IEEEmembership{Student~Member,~IEEE,}
        Robert~J.~Thomas,~\IEEEmembership{Life~Fellow,~IEEE,}
        and~Lang~Tong,~\IEEEmembership{Fellow,~IEEE}% <-this % stops a space
\thanks{The authors are with the School of Electrical and Computer Engineering, Cornell University, Ithaca, NY 14853, USA (e-mail: yj246@cornell.edu; rjt1@cornell.edu; ltong@ece.cornell.edu).}
\thanks{This work is supported in part by the U.S. Department of Energy Office of Electricity Delivery and Energy Reliability Consortium for Electric Reliability Technology Solutions (CERTS) program and the National Science Foundation under Grant CNS-1135844. Part of this work appeared in \cite{Ji15HICSS}.}}

\maketitle

\begin{abstract}
The short-term forecasting of real-time locational marginal price (LMP) and network congestion is considered from a system operator perspective. A new probabilistic forecasting technique is proposed based on a multiparametric programming formulation that partitions the uncertainty parameter space into critical regions from which the conditional probability distribution of the real-time LMP/congestion is obtained. The proposed method incorporates load/generation forecast, time varying operation constraints, and contingency models. By shifting the computation cost associated with multiparametric programs offline, the online computation cost is significantly reduced. { An online simulation technique by generating critical regions dynamically is also proposed, which results in several orders of magnitude improvement in the computational cost over standard Monte Carlo methods.}
\end{abstract}


\IEEEpeerreviewmaketitle

\section{Introduction}
\label{sec:intro}
As more renewable resources are integrated into the power system, and the transmission system operates closer to its capacity, congestion conditions become less predictable and locational marginal prices (LMP) more volatile. The increased congestion and LMP uncertainties pose significant challenges to the operator and market participants, which motivates us to consider the problem of short-term forecasting real-time LMP/congestion in the presence of generation, demand, and operation uncertainties.

The benefit of accurate LMP and congestion forecasts is twofold. For market participants, accurate forecast of real-time prices is valuable in risk management, bidding strategy development, and demand side participation.  The forecast prices  allow  market participants to make adjustments in advance to ensure competitive transactions. For system operators, on the other hand, forecast of transmission congestion is important for congestion management and system planning. European transmission system operators, for instance, use Intraday Congestion Forecast (IDCF) to improve real-time security assessment\cite{AUXENFANS14}. Intuitively, an LMP forecast should elicit generation participation at times of potential shortage thus alleviating future congestions.  Similarly, an LMP forecast can be used for demand response that results in shifting part of the load from peak to valley. An example of applying LMP forecast for inter-regional interchange is presented in \cite{ji2015scts}.

Currently, some system operators are providing real-time price forecasts. The Electric Reliability Council of Texas (ERCOT)\cite{ercot} offers a 1-hour ahead real-time LMP forecast, updated every 5 minutes. The Alberta Electric System Operator (AESO) \cite{aeso} provides two short-term price forecasts with prediction horizons of 2 hours and 6 hours, respectively.

Most LMP forecasting schemes provide only \textit{point forecast}, which gives a single quantity as the prediction.  For systems with high levels of uncertainty, a point forecast is rarely accurate, and impacts of prediction error are difficult to quantify.  A more attractive alternative is a\textit{ probabilistic forecast} that provides a full characterization of the LMP distribution.

Significant technical challenges exist for probabilistic forecasting real-time LMP and congestion.  First, reasonably accurate models for real-time dispatch and LMP are needed.  Second, network operating conditions and uncertainties need to be incorporated in real time. Finally, the forecasting algorithm needs to be simple and scalable for sufficiently large systems.

These challenges are daunting for external market participants who do not have access to network operating conditions and confidential information on bids and offers that influence LMPs.  On the other hand, if it is the system operator providing the forecast, as in the case of ERCOT or AESO, the barrier to efficient and accurate forecast is lowered.

\subsection{Summary of Contributions}
In this paper, we consider the real-time LMP and congestion forecasting problem from an operator perspective.  We focus on {\em probabilistic forecasting}  that, at time $t$, provides the conditional probability distribution  at time $t+T$ of the LMP vector and associated congestion,  given the system state at time $t$.  Here $T$ is referred to as the prediction horizon which is considered in the range of $T$ from 1 to 6 hours for short-term forecasts.

The key idea behind the proposed approach is the use of multiparametric programming that partitions the uncertainty space into critical regions with each region attached to a unique LMP and congestion pattern.  Thus, the problem of probabilistic forecasting reduces to computing the probabilities of random parameters falling in the set of critical regions.  When loads or generations (treated as negative loads) are random, their forecasts are incorporated to generate probabilistic LMP and congestion forecasts.

The proposed technique also provides several new features not present in existing methods.  For example, it can incorporate system contingency models and allow system constraints to vary with time.   The latter feature is relevant because network topology and thermal limits may be changed in real time by the system operator depending on operating conditions. In terms of the generation cost, the proposal can be applied to a linear (or piece-wise affine) function and a quadratic function. Computationally, the proposed method shifts a majority of the computation offline, which significantly reduces online computation cost.

{ An alternative algorithm that dynamically generates critical regions is also proposed. Because load and stochastic generation processes are physical processes, they are bounded and tend to concentrate around the mean trajectory. Their realizations thus only fall in a few critical regions instead of all over the entire parameter space.  By generating the critical regions that contain such realizations, the computation cost is reduced by several orders of magnitude comparing with standard Monte Carlo techniques.}
\subsection{Related Work}
{Much of the existing work deals with point forecasts of LMP by market participants who do not have access to real-time operating conditions and confidential offers and bids. For these techniques, historical data on LMP, load, and congestions drive the forecasting engine.  Literature on these techniques abounds. See \cite{Weron14EPF} and references therein. For probabilistic forecasting techniques by market participants, see approaches in Global Energy Forecasting Competition 2014 \cite{GEFCOM14}.}

There are several prior studies on LMP/congestion forecasting from the system operator perspective. The proposed technique in \cite{Min08} employs an online Monte Carlo sampling technique that, for each Monte Carlo sample path, solves an optimal power flow (OPF) problem, which is computationally expensive. Monte Carlo technique was also used in \cite{Ji13} where a reduction of the random variable dimension is made using a nonhomogeneous Markov chain model based on a partition of the system state space.

A particularly relevant prior work is \cite{Zhou11TPS} where the authors consider the problem of LMP/congestion forecasting from the vantage point of an external observer who has access to publicly available historical data only. Our work, in contrast, considers the forecasting problem from the vantage point of a system operator who has access to the system operating condition at the time of forecasting. In terms of forecasting methodology, the main difference between our approach and that in \cite{Zhou11TPS} lies in the different uses of conditioning in evaluating the conditional probability distribution of LMP/congestion.

The authors of \cite{Zhou11TPS} introduce and exploit the decomposition of a multi-dimensional load space into critical regions (called system pattern regions) that are estimated using historical data\footnote{The estimated critical regions are therefore random quantities.}. The work in \cite{Zhou11TPS} aims to address the following issue: Given a possible future point $L$ in a multi-dimensional load space, what is the probability distribution of the estimated critical regions that contain $L$? { Since each critical region corresponds to a specific LMP/congestion, the technique in \cite{Zhou11TPS} gives a heuristic estimate of the probability distribution of LMP/congestion by conditioning on load $L$ at some future point in time. }

In contrast to \cite{Zhou11TPS}, our objective is to forecast directly the probability distribution of future LMP/congestion, \textit{conditional on the current system operating point}. Because a system operator has access to all private and public information about system conditions, the critical regions are computed exactly via a multiparametric program. This allows us to incorporate load and generation forecasts and obtain the (conditional) probability distribution of future LMP/congestion directly.

Several techniques have been proposed to approximate the LMP distribution at a future time. In \cite{Bo09TPS}, a probabilistic LMP forecasting approach was proposed based on attaching a Gaussian distribution to a point estimate.  The advantage is that the technique can incorporate various point forecasting methods.  The disadvantage, on the other hand, is that network effects are not easy to incorporate.  The authors of \cite{mokhtari2009new} and \cite{davari2008determination}  approximate the probabilistic distribution of LMP using higher order moments and cumulants.  These methods are based on representing the probability distribution as an infinite series involving moments or cumulants. In practice, computing or estimating higher order moments and cumulants are very difficult; lower order approximations are necessary.

\vspace{-1em}\subsection{Organization and Notations}
This paper is organized as follows. Section \ref{sec:ed} introduces a model of the real-time economic dispatch and the ex-ante LMP computation. The key (known) results in multiparametric programming that form the basis of the proposed probabilistic forecasting approach are discussed in Section \ref{sec:mlp}. Details of the proposed techniques are given in Section \ref{sec:alg} and Section \ref{sec:dis}. Numerical results are presented in Section \ref{sec:sim} and followed by some concluding remarks in Section \ref{sec:conclusion}.

The notations used in this paper are standard.  For a random variable $x$, its expected value is denoted by $\bar{x}$.  We use  $\hat{\theta}$ to denote an estimate of parameter $\theta$. For a random process $y_t$, the forecast of $y_{t+T}$ using all the information available at time $t$ is denoted by ${y}_{t+T|t}$, where $T$ is the prediction horizon. We use $\bar{y}_{t+T|t}$ to denote the conditional mean of $y_{t+T}$ given $y_t$ (and possibly additional information at time $t$). The notation $z \sim \mathcal{N}(\mu,\Sigma)$ means that $z$ is a Gaussian random vector with mean $\mu$ and covariance matrix $\Sigma$. The indicator function $\mathbb{I}_{\mathcal{A}}(x)$ is one if $x\in\mathcal{A}$ and zero otherwise.

Vectors and matrices are in the real field.  $I_n$ denotes an $n\times n$ identity matrix and ${\bf 1}$ an $n$ dimensional column vector with all ones, where $n$ is the number of buses in the system. ${\bf 0}$  denotes a matrix/vector of zeros with different but compatible dimensions. We use superscript ``$^\intercal$'' to denote the transpose of a matrix and the complement of a set $\mathscr{I}$ is denoted by $\mathscr{I}^c$.

\section{Real-Time Economic Dispatch and Ex-Ante LMP}
\label{sec:ed}
We consider an ex-ante LMP model that arises from a real-time economic dispatch.  Our model is adopted from the stylized model in \cite{Ott03TPS} that captures the process of computing LMP by independent system operators. Specifically, the system operator solves a DC-OPF problem for optimal economic generation adjustment that meets load and stochastic generation forecasts for the next dispatch interval and satisfies generation and transmission constraints.

We describe here a simplified real-time economic dispatch formulation for the ex-ante LMP\footnote{By ex-ante LMP we mean that the computation of real-time dispatch and associated prices at time $t+1$ is based on the estimated system operating point at time $t$ and 1-step ahead load and generation forecasts.} model. For simplicity, we assume that each bus has a traditional generator, a stochastic generating unit and a load. The DC-OPF problem at time $t$ is given by
\begin{equation}\label{dcopf}%\vspace{-.25em}
\setlength\arraycolsep{2pt}\begin{array}{r l l}
\min\limits_g& c^\intercal g & \\
\text{subject to}&&\\
 & \textbf{1}^\intercal(g+w_{t+1|t}-d_{t+1|t})=0 &(\lambda_{t+1})\\
& L^- \leq {S} (g+w_{t+1|t}-d_{t+1|t}) \leq L^+ &(\mu_{t+1}^+ {\mu}_{t+1}^-) \\
 & G^-\leq g \leq G^+& \\
\end{array}%\hspace{-0.5em}
\end{equation}
where\\
\begin{tabular}{p{1.2cm} p{7cm}}
$c$& vector of real-time generation offers;\\
$g$& vector of ex-ante dispatch at time $t+1$;\\
$d_{t+1|t}$& vector of 1-step ahead load forecast at time $t$;\\
$w_{t+1|t}$& vector of 1-step ahead forecast of stochastic generation at time $t$;\\
${S}$& shift factor matrix;\\
$G^+/G^-$& vector of max/min generation capacities;\\
$L^+/L^-$& vector of max/min transmission limits;\\
$\lambda_{t+1}$& shadow price for the energy balance constraint  at time $t+1$;\\
$\mu_{t+1}^+/\mu_{t+1}^-$& shadow prices associated with max/min transmission constraints at time $t+1$.\\
\end{tabular}
The stochastic generation referenced above can be of any form of renewable integrations including renewable energy generation, distributed generation, and demand response. Here we assume that such stochastic generation is non-dispatchable with possible curtailment. The proposed forecasting method applies to other cost functions such as a piece-wise affine function or a quadratic function. This will be further addressed in the following sections

The LMP value at each bus is the sum of the marginal price of generation at the reference bus and the marginal congestion price at the location associated with the active transmission constraints. By the Envelope Theorem, the ex-ante LMP $\pi_{t+1}$ at time $t+1$ is given by
\begin{equation}\label{eqn:lmp}%\vspace{-.5em}
\pi_{t+1}= \textbf{1}\lambda_{t+1}-{S}^\intercal\mu^+_{t+1}+{S}^\intercal\mu^-_{t+1},
\end{equation}
where $\lambda_{t+1}, \mu^+_{t+1}$ and $\mu^-_{t+1}$ are from (\ref{dcopf}) at time $t$.

From (\ref{eqn:lmp}), we note that the LMP is determined by the marginal generator through $\lambda$ and the congestion pattern through $\mu^+$ and $\mu^-$. By a congestion pattern we mean the set of transmission lines where power flows have reached their limits. The forecasting of congestion pattern is a sub-problem of the forecasting of LMP.

\section{Multiparametric Programming}
\label{sec:mlp}

A multiparametric program (MPP) is a mathematical program indexed by a vector of parameters. The multiparametric programming problem is to solve the mathematical program for all values of the parameter vector (or a parameter space of interest). Consequently, an MPP captures simultaneous variations of distributed parameters in a network, which motivates its application in the problem of LMP forecast. Here we summarize some of the key theoretical results essential to the development of the proposed probabilistic forecast method.  The definitions and results follow \cite{Borrelli03}. In particular, we are interested in the linear and quadratic programs for which the multiparametric programming problems are called multiparametric linear programs (MPLP) and multiparametric quadratic programs (MPQP) respectively.

Consider a general right-hand side\footnote{By right-hand side we mean the parameter vector $\theta$ is on the right-hand side of the constraint inequalities.} MPP as follows:
\begin{equation}\label{mplp}%\vspace{-.5em}
\begin{array}{c}
\min\limits_x z(x) \text{ subject to } Ax \leq b+ E\theta \quad (y)\\
\end{array}\end{equation}
where $x$ is the decision vector, $\theta$ the parameter vector, $z(\cdot)$ the cost function\footnote{In this paper, we only consider the cost functions that are linear, piece-wise affine or quadratic.}, $y$ the Lagrangian multiplier vector, and $A$, $E$, $b$ are coefficient matrix/vector with compatible dimensions.

An MPP solver finds the subset $\Theta$ of a given polyhedron set such that the mathematical program (\ref{mplp}) is feasible and determines the expression $x^*(\theta) \in \mathscr{X}^*(\theta)$ of the optimizer (or one of the optimizers if it is not unique), where $x^*(\theta) $ is the optimal solution and $\mathscr{X}^*(\theta)$ the set of optimal solutions of (\ref{mplp}) given $\theta$.

The multiparametric programming analysis builds on the concepts of optimal partition and critical region.
\begin{mydef} [\hspace{-.1mm}\cite{Borrelli03}]\label{opt_part}
Let $\mathscr{J}$ denotes the set of constraint indices in (\ref{mplp}). For any subset $\mathscr{I} \subseteq \mathscr{J}$, let $A_\mathscr{I}$ and $E_\mathscr{I}$ be the corresponding submatrices of $A$ and $E$, respectively, consisting of rows indexed by $\mathscr{I}$. An \textit{optimal partition} of the index set $\mathscr{J}$ associated with parameter $\theta \in \Theta$ is the partition $(\mathscr{I} (\theta), \mathscr{I}^c(\theta))$ where
\begin{equation*}\hspace{-.35em}
\begin{tabular}{l}
$\mathscr{I}(\theta) \triangleq \{i \in \mathscr{J} | A_i x^*(\theta)=b+E_i \theta, \text{ for all } x^*(\theta)\in\mathscr{X}^*(\theta)\},$\\
$\mathscr{I}^c(\theta) \triangleq \{i \in \mathscr{J} | A_i x^*(\theta)<b+E_i \theta, \text{ for some } x^*(\theta)\in\mathscr{X}^*(\theta)\}.$\\
\end{tabular}%\vspace{-.5em}
\end{equation*}
For a given $\theta_0 \in \Theta$, let $(\mathscr{I}_0, \mathscr{I}^c_0) \triangleq (\mathscr{I}(\theta_0), \mathscr{I}^c(\theta_0))$. The \textit{critical region} related to the index set $\mathscr{I}_0$ is defined as
\begin{equation*}
\Theta_{\mathscr{I}_0} \triangleq \{ \theta \in \Theta |\mathscr{I}(\theta)=\mathscr{I}_0\},
\end{equation*}
which is the set of all parameters $\theta \in \Theta$ with the same active constraint set $\mathscr{I}_0$ at the optimum(s) of problem (\ref{mplp}).
\end{mydef}

By Definition \ref{opt_part}, the optimal partition specifies two sets of constraints by which the congestion pattern is determined: one set is a combination of active constraints at the optimum(s) and the other inactive. A critical region is the set of all parameter vectors for a certain optimal partition. The structure of critical region is a key property for the development of the online simulation technique, so we present it in Section \ref{sec:dis}. Here, we summarize the global property of critical region and its connection with the solution structure of MPP (\ref{mplp}). The connection between the feasible parameter space and critical regions is summarized in the lemma below.
%\vspace{-.25em}
\begin{lem}[\hspace{-.1mm}\cite{Borrelli03}]\label{lem:partition}
The feasible parameter space $\Theta$ can be partitioned into critical regions $\{\Theta_i\}$. If there is no (primal or dual) degeneracy, such a partition is unique.
\end{lem}%\vspace{-.25em}
Assume that there is no (primal or dual) degeneracy. In the linear (or piece-wise affine) cost case, the optimizer is unique and the associated Lagrangian multiplier vectors are constant, for all parameters within each critical region. Therefore, each critical region is associated with a unique LMP vector. In the quadratic cost case, the LMP vector is a uniquely defined affine function of the parameter within each critical region, which follows the theorem below.
%\vspace{-.25em}
\begin{thm}[\hspace{-.1mm}\cite{bemporad2002explicit}]\label{thm:quad}
If the MPQP problem (\ref{mplp}) is not degenerate\footnote{Note that the cost function of the MPQP problem is assumed to be strictly convex.}, the Lagrangian multiplier vector associated with the optimal solution $x^*(\theta)$ is a uniquely defined affine function of $\theta$ within each critical region.
\end{thm}%\vspace{-.25em}
Note that degeneracy may happen for sufficiently large systems or high dimensional parameters. Although the Lagrangian multiplier vector may not be unique in such cases, a consistent tie breaking rule can be introduced to guarantee the uniqueness.

A key property of the optimizer for both MPLP and MPQP is as follows.
\vspace{-.25em}\begin{thm}[\hspace{-.1mm}\cite{Borrelli03}\cite{bemporad2002explicit}]\label{thm:affine}
If the optimizer $x^*(\theta)$ of MPLP/MPQP (\ref{mplp}) is unique, then it is continuous and piece-wise affine over $\Theta$. In particular,  $x^*(\theta)$ is an affine function of $\theta$ in each critical region $\Theta_i$. If there exists multiple optimizers, it is always possible to define a continuous and piece-wise affine function $x^*(\theta)\in\mathscr{X}^*(\theta)$ for all $\theta\in\Theta$.
\end{thm}\vspace{-.25em}
This theorem is crucial to the development of the proposed forecasting technique. In essence, we treat the randomness as a vector of parameter and solve the MPPs ahead of time. The online computation of the optimal solution is highly simplified because the piece-wise affine mapping between the optimal solution and the parameter vector has already been obtained offline.

The problem of solving MPPs, including the computation of critical regions and corresponding piece-wise affine functions,  has been well studied. The first method for solving right-hand side MPLPs was proposed by Gal and Nedoma \cite{Gal72}. A geometric approach for critical region partition was described in \cite{Borrelli03}. The piece-wise affine relationship of the parameter vector and the optimal solution was proved for MPLP and MPQP in \cite{Gal72} and \cite{bemporad2002explicit}, respectively. In this paper, all calculations related to MPPs are performed using MPT3 toolbox\cite{MPT3} except for dynamical generations of critical regions in the proposed online simulation technique presented in Section \ref{sec:dis}.

\section{Probabilistic Forecasting Algorithm}\label{sec:alg}
\emph{Probabilistic forecasting}, in contrast to point forecasting, aims to provide the probability distribution of a future LMP. In particular, given the estimated system operating point  at time $t$ and load and generation forecasts, the probabilistic forecast  at time $t$ of the LMP at time $t+T$ is given by the conditional probability distribution $f_{t+T|t}$ of the LMP vector. Entries of the LMP vector are LMPs at individual buses in the system.  Since each congestion pattern can be mapped from an LMP vector, we only discuss the probabilistic forecasting of LMP here; the probability distribution of congestion can be obtained from that of LMP.

The key to probabilistic LMP forecasting is to capture spatial and temporal dependencies.  Spatial correlations among LMPs arise naturally from the optimization that governs the real-time dispatch.  Temporal correlations, on the other hand, are the results of time dependencies in load/generation forecasts. The system randomness may also include occurrences of random contingencies. In this section, we first give an overview of the proposed forecasting algorithm. Details on addressing these dependencies are then discussed.

\subsection{Overview} \label{subsec:over}
\begin{figure}\begin{psfrags}
\psfrag{C0}[l]{\footnotesize $\Theta_7$}
\psfrag{C1}[c]{\footnotesize $\Theta_1$}
\psfrag{C2}[c]{\footnotesize $\Theta_2$}
\psfrag{C3}[c]{\footnotesize $\Theta_3$}
\psfrag{C4}[c]{\footnotesize $\Theta_4$}
\psfrag{C5}[c]{\footnotesize $\Theta_5$}
\psfrag{C6}[c]{\footnotesize $\Theta_6$}
\psfrag{D}[c]{\small Feasible Parameter}
\psfrag{F}[c]{\small Space $\Theta$}
\psfrag{pi}[l]{\small $\theta_{t+T}$}
\psfrag{pi0}[c]{\small $\theta_{t}$}
\psfrag{t}[c]{\tiny $t$}
\psfrag{t1}[c]{\tiny $t+1$}
\psfrag{ptT}[c]{\small ${f}_{t+T|t}$}
\psfrag{pt}[c]{\small $f_t$}
\includegraphics[width=0.5\textwidth]{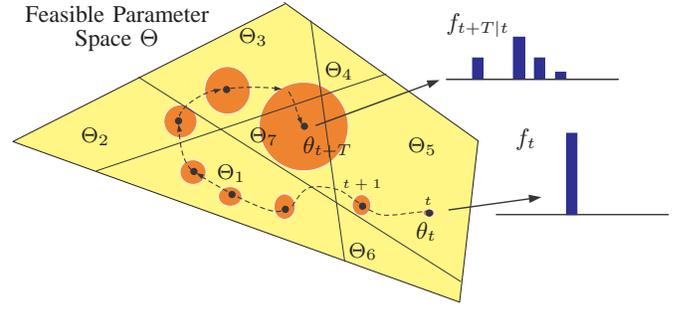}
\caption{Geometric intuition of the proposed algorithm.}\label{fig:geo}
\end{psfrags}\end{figure}

The basic idea of the proposed probabilistic forecasting technique is using multiparametric programming analysis to characterize the variation of the real-time LMP with respect to the random load and generation. By formulating the DC-OPF problem (\ref{dcopf}) as an MPP in the form of (\ref{mplp}), the real-time LMP can be expressed as a function of the random load and generation. The distribution of the LMP vector at a future time can be obtained from the probabilistic forecasts of the stochastic load and generation.

As illustrated in Fig. \ref{fig:geo}, load and stochastic generation forecasts in (\ref{dcopf}) are treated as parameters, denoted by $\theta=(d,-w)$, where $d$ is the vector of parametric load and $w$ the vector of parametric generation (which is treated as a negative load).  The feasible parameter space $\Theta$ is partitioned into critical regions $\{\Theta_1, \cdots, \Theta_7\}$. Within each region $\Theta_i$, the optimal dispatch is associated with the same Lagrange multipliers, and hence a unique LMP vector $\pi_i$ for all $\theta \in \Theta_i$. Given the network parameters, the MPLP solver computes the partition $\{\Theta_i\}$. Correspondences $\{\Theta_i, \pi_i\}$ are then obtained by the Lagrange multipliers and the LMP model (\ref{eqn:lmp}).  Note that this computation does not depend on the actual realization of load and generation. Therefore, the computation of the partition and the correspondences can be obtained offline.

Consider now the trajectory of a realization of the random load and generation process $\theta_t$ as illustrated in Fig. \ref{fig:geo}.  Given the realization $\theta_t$ and system measurements at time $t$, we are interested in the conditional probability distribution\footnote{This is the case for linear or piece-wise affine cost functions. If the cost function in the real-time economic dispatch is quadratic, the forecast distribution $f_{t+T|t}$ of LMP $\pi_{t+T}$ can be obtained from the (continuous) distribution of $\theta_{t+T}$, and the uniquely defined affine function of the Lagrangian multiplier vector within each critical region.}
\begin{equation}\label{prob_gen}%\vspace{-.2em}
{f}_{t+T|t}(i)=\mathbb{P}[\theta_{t+T}\in \Theta_i|\theta_{t}].
\end{equation}

As depicted by the shaded circles in Fig. \ref{fig:geo}, uncertainties associated with load and generation forecasts increase with time. At time $t$, the realization of parameter $\theta_{t}$ is known and thus the distribution $f_t$ is an unit vector. But parameter $\theta_{t+T}$ may take values from several critical regions where LMP values and congestion patterns are different. Therefore, the forecast probability mass function ${f}_{t+T|t}$ of $\theta_{t+T|t}$ may have several non-zero elements.

The proposed probabilistic forecasting algorithm involves two parts explained in the following subsections: the computation of critical regions and the estimation of conditional probability distributions, where the former is computed offline and the latter online.
\subsection{Forecast with Varying Operational Conditions}\label{subsec:base}
We describe here a baseline formulation from which critical regions are obtained. We formulate the DC-OPF (\ref{dcopf}) used to compute LMP at time $t+T$ as the following right-hand side MPLP with the uncertainty parameter $\theta$ consisting of only stochastic load and generation.
\begin{equation}\label{dcopf2}
\begin{array}{l}
\hspace{.38in}\min\limits_g\;  c ^\intercal g \\
\text{ subject to } \\
\left[ \begin{smallmatrix}
\textbf{1}^\intercal\\
-\textbf{1}^\intercal\\
{S}_{t+T-1} \\
-S_{t+T-1} \\
I_n\\
-I_n\\
\end{smallmatrix}\right]
g\leq
\left[\begin{smallmatrix}
\textbf{1}^\intercal & \textbf{1}^\intercal\\
-\textbf{1}^\intercal & -\textbf{1}^\intercal\\
{S}_{t+T-1} & {S}_{t+T-1}\\
-{S}_{t+T-1} & -{S}_{t+T-1}\\
\textbf{0}&\textbf{0}\\
\textbf{0}&\textbf{0}\\
\end{smallmatrix}\right]\theta
+
\left[\begin{smallmatrix}
0\\0\\ L^+_{t+T-1}\\ -L^-_{t+T-1}\\ G^+_{t+T-1}\\ -G^-_{t+T-1}
\end{smallmatrix}\right]\\
\end{array}\hspace{-0.5em}
\end{equation}
where $S_{t+T-1}$, $L^+_{t+T-1}$, $L^-_{t+T-1}$, ${G}^+_{t+T-1}$, and ${G}^-_{t+T-1}$ are shift factor matrix, vectors of max and min transmission limits, and vectors of max and min generation capacities at time $t+T-1$, respectively.

Here we allow {\em time varying but  known system parameters} such as shift factors, flow limits on transmission lines, and limits on generations, restricting uncertainties only to load and generation.  The idea is to include deterministically scheduled events in the forecasting problem.  Examples include the scheduled changes in network topology \cite{hedman2011review}, and generation capacity and transmission limit \cite{NE19}.
\subsection{ Forecast in the Presence of Probabilistic Contingencies}

The baseline MPLP formulation described in Section \ref{subsec:base} can be extended to include the presence of probabilistic contingencies for unexpected events. For example, a transmission line may be tripped in a storm or generation capacity reduced due to faults.  Such uncertainties in the system parameters need to be handled differently from those associated with the stochastic load and generation.

Because unexpected changes of system configurations are typically small probability events, we assume that there are a total of $K$ possible system configurations at time $t+T$.  From historical data, we assume that system configuration $k$ happens with an estimated probability $\hat{p}_k$.

We solve the baseline MPLP (\ref{dcopf2}) for each system configuration and obtain critical regions for system configuration $k$ denoted by $\{\Theta_i^{(k)}\}$.  By the total probability theorem, the probabilistic forecast of LMP at time $t+T$ is therefore given by\vspace{-.5em}
\begin{equation}\vspace{-.5em}
f_{t+T|t} = \sum_{k=1}^K  \hat{p}_k f_{t+T|t}^{(k)},
\end{equation}
where $f^{(k)}_{t+T|t}$ is the forecast distribution under system configuration $k$ using critical regions $\{\Theta_i^{(k)}\}$ for $k=1,2, \cdots, K$.

We illustrate the above idea with an example. Consider the case when at most one of $K$ contingencies can occur between time $t$ and $t+T$.  We model the probabilistic contingency as independent tosses of a $K+1$ faced dice where contingency $k$ occurs with probability $p_k$ and no contingency with $p_0=1-\sum_{k=1}^K p_k$.   We further assume that, once a particular contingency occurs, it remains until time $t+T$.
We then solve each MPLP problem (\ref{dcopf2}) for all $K+1$ possible system configurations, including the normal condition. The probabilistic forecast of LMP at time $t+T$ is given by
\begin{equation}
f_{t+T|t}= \left\{
\begin{array}{l l}
f_{t+T|t}^{(k)} &\text{if contingency $k$ occurs}\\
\sum_{k=0}^K {p}_k f_{t+T|t}^{(k)} & \text{otherwise}\\
\end{array} \right.
\end{equation}
where $f_{t+T|t}^{(k)}$ is the forecast distribution under system configuration $k$ with the feasible space $\Theta_{t+T|t}^{(k)}$ for $k=0,1,\cdots, K$. Note that system configuration $0$ denotes the normal condition.

Having obtained the critical regions, we now consider the problem of computing the conditional distribution of LMP at time $t+T$, given the load and generation forecast at time $t$.

\subsection{Probability Distribution Estimation}\label{sec:est}
The estimation of conditional probabilities in (\ref{prob_gen}) depends on statistical models of load and generation. Such models can be obtained from either models of the load and stochastic generation process or specific prediction methods used to generate load and stochastic generation forecasts.

As illustrations, we present a directional Gaussian random walk model and an autoregressive (AR) noise model for the random load and generation processes here. It should be noted that any statistical model or prediction method can be applied. The purpose of using these models is to gain insights into the behavior of forecasting performance by taking advantage of some of analytically tractable properties.

\subsubsection{A Directional Random Walk Model}We first consider a random walk model of the load/stochastic generation based on a given mean trajectory. Such a model represents a case of minimally informative forecast. Note that the mean trajectory can be any available forecast. For example, a reasonable mean trajectory is the day-ahead load forecast.

Assume that load/stochastic generation $\theta_{t}$ follows a random walk process with a (known) mean trajectory $\bar{\theta}_{t}$:
\begin{equation}\label{eqn:rw}
\theta_{t}=\theta_{t-1}+\bar{\theta}_{t}-\bar{\theta}_{t-1}+\epsilon_{t},
\end{equation}
where $\epsilon_t \sim \mathcal{N}(\mathbf{0},\Sigma)$.
Given the realization $\theta_t$ at time $t$, the actual load/generation at time $t+T$ is given by%\vspace{-.5em}
\begin{equation}%\vspace{-.5em}
\theta_{t+T}=\theta_t+\bar{\theta}_{t+T}-\bar{\theta}_t+\sum_{i=t+1}^{t+T}\epsilon_{i}.
\end{equation}

Therefore, the distribution of $\theta_{t+T}$ conditioning on $\theta_t$ is:
\begin{equation}\label{dist:rw}
\theta_{t+T}\sim\mathcal{N}(\bar{\theta}_{t+T|t}, \Sigma_T)
\end{equation}
 where $\bar{\theta}_{t+T|t}=\theta_t+\bar{\theta}_{t+T}-\bar{\theta}_t$ is the conditional mean of $\theta_{t+T}$ and $\Sigma_T=T\Sigma$ is the cumulative variance within prediction horizon $T$.

\subsubsection{An AR Noise Model}The second model we consider is an AR(1) noise model where we assume the deviation of the load or generation from the expected value is an AR(1) process. This is a case when the load or stochastic generation is highly structured. In particular,
\begin{equation}\label{eqn:ar}%\vspace{-.5em}
\theta_t=\bar{\theta}_t+a_t, a_t=\phi a_{t-1}+\epsilon_t,
\end{equation}
where $\bar{\theta}_t$ is the (known) mean trajectory, $\phi$ the parameter of the AR process, and $\epsilon_t \sim \mathcal{N}(\mathbf{0},\Sigma)$.
Given the realization $\theta_t=\bar{\theta}_t+a_t$ at time $t$, the noise at time $t+T$ is given by
\begin{equation}%\vspace{-.5em}
a_{t+T}=\phi^T(\theta_t-\bar{\theta}_t)+\sum_{i=0}^{T-1}\phi^{i}\epsilon_{t+T-i},
\end{equation}
and the actual load/generation at time $t+T$ is given by%\vspace{-.5em}
\begin{equation}%\vspace{-.5em}
\theta_{t+T}=\bar{\theta}_{t+T}+\phi^T(\theta_t-\bar{\theta}_t)+\sum_{i=0}^{T-1}\phi^{i}\epsilon_{t+T-i}.
\end{equation}

Therefore, the distribution of $\theta_{t+T}$ conditioning on $\theta_t$ is:
\begin{equation}\label{dist:ar}
\theta_{t+T}\sim\mathcal{N}(\bar{\theta}_{t+T|t}, \Sigma_T)
\end{equation}
where $\bar{\theta}_{t+T|t}=\bar{\theta}_{t+T}+\phi^T(\theta_t-\bar{\theta}_t)$ is the conditional mean of $\theta_{t+T}$,  and $\Sigma_T=\sum_{i=0}^{T-1}\phi^{i}\Sigma$ the cumulative variance within prediction horizon $T$.

To sum up, for both models, the conditional probability of $\theta_{t+T}$ falling in critical region $\Theta_i$ given $\theta_t$ is:
\begin{equation}\label{prob}
{f}_{t+T|t}(i)=\int_{\Theta_i}\frac{\exp\{ -\frac{1}{2}({x}-{\bar{\theta}_{t+T|t}})^\intercal\Sigma_T^{-1}({x}-{\bar{\theta}_{t+T|t}})\} }{\sqrt{(2\pi)^n|\Sigma_T|}}d{x},
\end{equation}
where $\bar{\theta}_{t+T|t}$ and $\Sigma_T$ are model associated statistics given in (\ref{dist:rw}) and (\ref{dist:ar}).

In general, Monte Carlo techniques are necessary to estimate the conditional probability (\ref{prob}). To accelerate the sampling process, importance sampling technique is used. In particular, for each critical region $\Theta_i$, instead of drawing values from distribution $\mathcal{N}(\bar{\theta}_{t+T|t}, \Sigma_T)$, we use $\mathcal{N}(\bar{v}(\Theta_i), \Sigma_T)$ where $\bar{v}(\Theta_i)$ is the mean of all vertices of critical region $\Theta_i$. The estimate of $f_{t+T|t}(i)$ is then given by%\vspace{-.5em}
\begin{equation}\label{eqn:is}%\vspace{-.5em}
 \hat{f}_{t+T|t}(i)=\frac{1}{N}\sum_{j=1}^N\frac{\mathbb{I}_{\Theta_i}(s_j)g(s_j)}{h(s_j)},
\end{equation}
where samples $\{s_1, \cdots, s_N\}$ are drawn from $\mathcal{N}(\bar{v}(\Theta_i), \Sigma_T)$, and $g(\cdot)$ and $h(\cdot)$ are probability density functions (PDFs) of distribution $\mathcal{N}(\bar{\theta}_{t+T|t}, \Sigma_T)$ and $\mathcal{N}(\bar{v}(\Theta_i), \Sigma_T)$, respectively. Note that the importance distribution $\mathcal{N}(\bar{v}(\Theta_i), \Sigma_T)$ only shifts the mean of the nominal distribution $\mathcal{N}(\bar{\theta}_{t+T|t}, \Sigma_T)$ but keeps the variance the same.

\subsubsection{A Quadratic Cost Case}
If the cost function in the DC-OPF (\ref{dcopf}) is quadratic, the distribution of the LMP $\pi_{t+T}$ cannot be estimated by the conditional probabilities of $\theta_{t+T}$ falling in each critical region. Because the LMP in this case is an affine function of the parameter vector by Theorem \ref{thm:quad}.

Here we derive the conditional distribution $f_{t+T|t}$ at time $t$ of the future LMP $\pi_{t+T}$ at time $t+T$ given the conditional distribution of $\theta_{t+T}$.  By Theorem \ref{thm:quad} and the LMP formulation (\ref{eqn:lmp}), for each critical region $\Theta_i$, there exists an affine function $\pi_i(\cdot): \Theta_i\rightarrow\Pi_i$ such that
\[\pi_i(\theta) =U_i\theta+v_i, \text{ for all }\theta \in \Theta_i,\]
where $\Pi_i$ is the codomain and $U_i$ and $v_i$ are associated coefficient matrix/vector. Note that these coefficients can be obtained from the affine function of the Lagrangian multiplier vector and the LMP formulation (\ref{eqn:lmp}).

Given the conditional distribution $\mathcal{N}(\bar{\theta}_{t+T|t}, \Sigma_T)$  of $\theta_{t+T}$, the conditional PDF of $\pi_{t+T}$ is given by%\vspace{-.5em}
\begin{equation} \label{dist:pi}%\vspace{-.5em}
f_{t+T|t}(\pi)=\sum_{i=1}^N \mathbb{I}_{\Pi_i}(\pi)f_{t+T|t,\Pi_i}(\pi),
\end{equation}
where $N$ is the number of critical regions, and $f_{t+T|t, \Pi_i}(\cdot)$ the PDF of $\mathcal{N}(U_i\bar{\theta}_{t+T|t}+v_i, U_i^\intercal \Sigma_TU_i)$, which is the conditional distribution of $\pi_{t+T}$ in codomain $\Pi_i$.

In summary, the proposed algorithm treats load/generation as parameters, formulates the DC-OPF (\ref{dcopf}) as an MPP (\ref{mplp}), determines the critical regions, and computes the conditional distribution of LMP and congestion using load/generation forecasts and the real-time operation conditions. These system conditions, such as transmission rate, generation capacity, and network topology, are allowed to vary with time but known to the operator at the time of forecast. Contingency models are also incorporated in the proposed technique.
{\section{Online Forecasting via Dynamic Critical Region Generation}\label{sec:dis}
A limiting factor in the proposed technique above is the computational cost associated with the multiparametric programming. Since the solution structure is characterized by critical regions, all critical regions that partition the parameter space have to be calculated. Although such computation can be made offline, it may not be computationally tractable for large systems because the number of critical regions may grow exponentially with the number of constraints.

In this section, we propose an online Monte Carlo technique, referred to as dynamic critical region generation (DCRG),  that significantly reduces the computation cost. The  idea is to take advantage of the fact that, in practice, random load and generation processes are bounded and tend to concentrate around their mean trajectories. As a result, a small fraction of critical regions represents the overwhelming majority of observed critical regions.  When a parameter falls in a critical region that was visited before, the Lagrange multipliers that are used to generate LMPs can be obtained directly from the affine mappings associated with that critical region, without solving the DC-OPF (\ref{dcopf}).

The key idea of DCRG, therefore, is to compute on demand the critical region and the associated coefficients of the affine mapping of parameter to the LMP.  This computation, fortunately, is no more than elementary matrix inversions and multiplications.  The computation procedure is given by the following theorem that summarizes known results in MPLP/MPQP \cite{Borrelli03}\cite{bemporad2002explicit}. The proof is provided in the Appendix.

\begin{thm}\label{thm:mp_soln_struct}
Given a parameter $\theta_0$, let $(\mathscr{I}_0, \mathscr{I}^c_0)$ be the optimal partition of the index set $\mathscr{J}$ of (\ref{mplp}) associated with $\theta_0$. Let $A_{\mathscr{I}_0}, E_{\mathscr{I}_0}$ and $b_{\mathscr{I}_0}$ be, respectively, the submatrices of $A$, $E$ and subvector of $b$ corresponding to the index set $\mathscr{I}_0$.  Let $A_{\mathscr{I}^c_0}, E_{\mathscr{I}^c_0}$ and $b_{\mathscr{I}^c_0}$ be similarly defined for the index set $\mathscr{I}^c_0$.  Assume that MPP (\ref{mplp}) is neither primal nor dual degenerate for all $\theta$. Denote the critical region that contains $\theta_0$ by $\Theta_{\mathscr{I}_0}$.
 \begin{enumerate}
 \item[(1)] For the MPLP (\ref{mplp}) with  cost function $z(x)=c^\intercal x$, the critical region $\Theta_{\mathscr{I}_0}$ is given by
  \begin{equation}\label{eqn:mlpcr}
  \Theta_{\mathscr{I}_0}=\left\{\theta \big| A_{\mathscr{I}^c_0}A_{\mathscr{I}_0}^{-1}(b_{\mathscr{I}_0}+E_{\mathscr{I}_0}\theta)<b_{\mathscr{I}^c_0}+E_{\mathscr{I}^c_0}\theta\right\}.\end{equation}
  \item [(2)] For the MPQP (\ref{mplp}) with  cost function $z(x)=\frac{1}{2}x^\intercal H x$ where $H$ is  positive definite,   the critical region $\Theta_{\mathscr{I}_0}$ is given by
   \begin{equation}\label{eqn:mqpcr} \Theta_{\mathscr{I}_0}=\{\theta| \theta\in \mathscr{P}_p  \bigcap \mathscr{P}_d \}\end{equation}
   where $\mathscr{P}_p$  and  $\mathscr{P}_d$ are polyhedra defined by
   \[\begin{array}{l}\mathscr{P}_p=\left\{\theta \left| \begin{array}{l}A_{\mathscr{I}^c_0}H^{-1}A_{\mathscr{I}_0}^\intercal(A_{\mathscr{I}_0}H^{-1}A_{\mathscr{I}_0}^\intercal)^{-1}(b_{\mathscr{I}_0}+E_{\mathscr{I}_0}\theta)\\
\qquad \qquad <b_{\mathscr{I}^c_0}+E_{\mathscr{I}^c_0}\theta \\ \end{array} \right.\right\}\\
\mathscr{P}_d=\{\theta|(A_{\mathscr{I}_0}H^{-1}A_{\mathscr{I}_0}^\intercal)^{-1}(b_{\mathscr{I}_0}+E_{\mathscr{I}_0}\theta)\leq \mathbf{0}\}.\end{array}\]
\end{enumerate} \end{thm}

In applying DCRG to LMP forecasting using online Monte Carlo simulation, we generate samples of random load or generation, either based on load/generation models or from historical data.  Instead of directly simulating the real-time market operation as in \cite{Min08}, we check if the generated sample falls into a critical region that has been used before.  If it does, the LMP can be generated using the affine mapping of Lagrangian multipliers directly without solving the DC-OPF (\ref{dcopf}).  If the parameter does not belong to any critical region in the database, a DC-OPF is solved and a critical region containing the parameter is computed according to (\ref{eqn:mlpcr}) for linear cost case or (\ref{eqn:mqpcr}) for quadratic cost case in Theorem \ref{thm:mp_soln_struct}.
}
\section{Evaluation}
\label{sec:sim}
In this section, we present simulation results to compare performances of the proposed probabilistic forecasting algorithm with some benchmark methods. We first show results of a 3 bus system with a linear cost function to gain insights into the behavior of the proposed algorithm under various scenarios. { Simulations using the IEEE 118 bus system with a quadratic cost function are then presented to demonstrate the scalability of the proposed algorithm and the effectiveness of the online heuristic approach given in Section \ref{sec:dis}.}
\subsection{Benchmarks and Performance Measure}
{ We compared the proposed techniques with some existing benchmarks for forecasting and computation performance.  Since, to our best knowledge, there is no probabilistic forecasting techniques for the operator in the literature, we used the direct Monte Carlo simulation method  proposed in \cite{Min08} as the probabilistic forecasting benchmark\footnote{We want to point out that the direct Monte Carlo simulation approach generates exactly the same probabilistic forecast as the proposed technique. }. We also included comparisons of the proposed technique with two well known point forecasting methods to illustrate the performance gain. In particular, the deterministic prediction uses the mean trajectory of load/stochastic generation $\bar{\theta}_{t+T}$ to calculate LMP and congestion pattern at time $t+T$. The certainty equivalence prediction incorporates measurements at time $t$ and uses the conditional mean trajectory $\bar{\theta}_{t+T|t}$.
}

Before presenting numerical results, we introduce a performance evaluation metric of probabilistic forecasts. The LMP\footnote{Note that the discreteness of LMP is only for linear or piece-wise affine cost functions.}/congestion pattern is a discrete random vector. The probabilistic forecast of such a random quantity belongs to the so-called categorical forecast, and its performance is measured by the consistency as well as the statistical concentration of the forecast.  A standard metric\cite{GneitingRaftery07ScoreRule} for this type of forecast is the Brier Score (BS) \cite{Brier} that measures the average distance (2-norm) between the forecast distribution ${f}_{t+T|t}$  and the point mass distribution at the realized random variable  $\pi_{t+T}$.  Specifically,
\begin{equation}\label{eqn:bs}%\vspace{-.25em}
\text{BS}({f}_{t+T|t})=\mathbb{E}\|{f}_{t+T|t}-\delta(\pi_{t+T})\|^2,
\end{equation}
where the expectation is taken over all randomness between time $t$ and $t+T$.  In (\ref{eqn:bs}), ${f}_{t+T|t}$ is the conditional probability vector whose $i$th entry is given by ${f}_{t+T|t}(i) = \Pr(\pi_{t+T}=\pi_i)$, and $\delta (x)$ is the unit vector that is one at entry $x$ and zero elsewhere. This score ranges from $0$ for a perfect forecast to $2$ for the worst possible forecast.

Since BS is a succinct formula to measure the overall performance in terms of uncertainty, reliability and resolution, we also provide a more intuitive assessment --- reliability diagram. Reliability diagram is a  graph of the observed frequency of an event plotted against the forecast probability of an event. It measures how closely the forecast probabilities of an event correspond to the actual chance of observing the event.
\subsection{Case Study: A 3 Bus System}
\begin{figure}\hspace{1em}\begin{psfrags}
\psfrag{F1}[l]{}
\psfrag{F2}[l]{}
\psfrag{F3}[l]{}
\psfrag{F4}[l]{}
\psfrag{F5}[l]{}
\psfrag{F6}[l]{}
\psfrag{C1}[l]{\tiny $c_1=10$ \$/MWh }
\psfrag{C2}[l]{\tiny $c_2=15$ \$/MWh}
\psfrag{G1}[l]{\tiny $G_1^+=130$ MW}
\psfrag{G2}[l]{\tiny $G_1^-=0$ MW}
\psfrag{G3}[l]{\tiny $G_2^+=200$ MW}
\psfrag{G4}[l]{\tiny $G_2^-=0$ MW}
\hspace{1em}\includegraphics[width=0.35\textwidth]{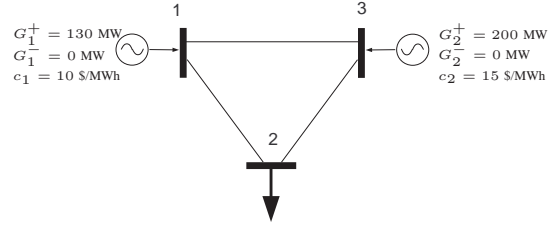}\vspace{-1em}
\caption{3 bus system.}\label{fig:3bus}%
\end{psfrags}\end{figure}

Consider a 3-bus system as depicted in Fig. \ref{fig:3bus}. Generator incremental costs and capacity limits are presented in the figure. All lines are identical with the maximum capacity of $100$ MW.

\subsubsection{Baseline}We first evaluated the baseline algorithm with the two load models described in Section \ref{sec:est}. Note that only the load at bus 2 was stochastic, which gave the one dimensional parametric linear program.

Table 1 shows the critical regions of the parametric linear program, and the associated LMPs and congestion patterns.  In this case, the parameter load $d$ at bus 2 was partitioned into three segments $\mathscr{D}=\{(0,130), (130, 170), (170,200)\}$.

\begin {table}\begin{center}\caption{Critical regions, LMPs and congestion patterns for the baseline\protect\footnotemark.}
\begin{tabular}{|c|c|c|c|}  \hline
&Critical Region&LMP & Congestion\\ \hline
1&(0, 130)& (10, 10, 10)& (0, 0, 0)  \\ \hline
2&(130, 170) & (15, 15, 15)& (0, 0, 0)\\ \hline
3&(170, 200)& (10, 20, 15) & (1, 0, 0) \\ \hline
\end{tabular}
\label{table:3bus_load}\vspace{-1em}
\end{center}\end {table}
\footnotetext{In Table \ref{table:3bus_load}, the triple for LMP contains price values at bus 1-3. For congestion, the triple summarizes status of line 1-2, line 1-3, and line 2-3, respectively, where ``$1$'' indicates positive congestion (the flow reaches the line limit in the positive direction), ``$-1$'' negative congestion, and ``$0$'' no congestion. }

\begin{figure}\centering
\subfloat[Random walk model]{\includegraphics[scale=0.6]{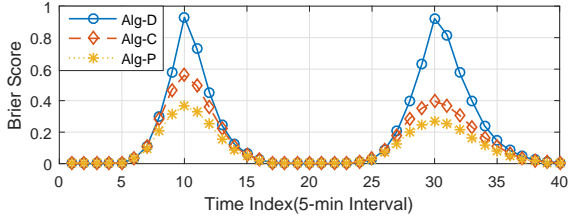}\label{err1}}\\ \vspace{-1em}
\subfloat[AR(1) noise model]{\includegraphics[scale=0.6]{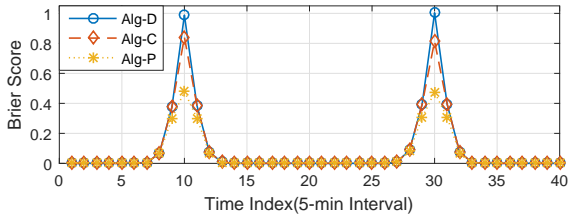}\label{err2}}
\caption{Impact of load statistical models. }\label{fig:load_model}%\vspace{-1em}
\end{figure}

We used a straight line ranging from $110$ MW to $190$ MW with $2$ MW increments as the mean trajectory of load $\bar{d}_t$. The coefficient $\phi$ in AR(1) noise model was set at $0.9$. The independent noise sequence $\epsilon_t$ in both models followed the standard normal distribution, \textit{i.e.}, $\mathcal{N}(0,1)$.  Monte Carlo simulations were used to obtain estimated BSs in Fig. \ref{fig:load_model}.

From Fig. \ref{fig:load_model}, we observed that the proposed probabilistic forecasting algorithm ``Alg-P'' consistently outperforms the deterministic ``Alg-D'' and certainty equivalence ``Alg-C'' predictors in both load models. The superior performance of the proposed technique in these two extreme models (a minimally informative model and a highly structured model) shows its capability of incorporating different load forecasting methods and its forecasting power. Two interesting phenomena are worth closer examinations.  First, for both load models, peaks occurred at the boundaries of two neighboring critical regions when the mean load $\bar{d}_t$ is $130$ at $t=10$ and $170$ at $t=30$.   At the boundary point, the probability of $d_{t+T}$ falling in either the left or the right critical region was the same. Roughly half of the time Alg-D predicted the LMP correctly, the other half was completely wrong.  From the definition in (\ref{eqn:bs}), the BS of Alg-D should be $1$.

Second, a more subtle point, the scores for the random walk model showed a slight asymmetry with respect to boundaries of neighboring critical regions: the BSs of  Alg-P and Alg-C at the second peak were lower than that of the first peak, and the ranges of non-zero score neighborhood of the second peak (from time $24$ to $39$) of all three algorithms were wider than that of the first peak (from time $5$ to $16$).  This phenomenon arose primarily from the process of generating the sample trajectory of load. Since the entire sample trajectory was generated at once, the deviation $|d_t-\bar{d}_t|$ from the mean grows over time which leads to a bigger variance of the time crossing the second boundary $170$ than that of  the time crossing the first boundary $130$. In other words, comparing to the probability of crossing the first boundary $130$ at $t=10$, the probability of crossing the second boundary $170$ at $t=30$ is lower, but the probability in its neighborhood is higher. Therefore, the scores of Alg-P and Alg-C at the second peak were lower, and the ranges of non-zero score neighborhood bigger.

\begin{figure}\centering
\includegraphics[scale=0.6]{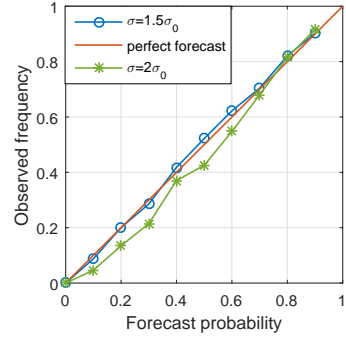}
\caption{Reliability diagrams for critical region 1 with different load forecast uncertainties.}\label{fig:reliability}%\vspace{-1em}
\end{figure}

To evaluate the robustness and reliability of the proposed algorithm, we tested different levels of load forecast uncertainty. Specifically, we varied the variance of the noise in the random walk load model: by default,  $\epsilon\sim\mathcal{N}(0,\sigma^2_0)$, where $\sigma_0^2=1$, and then we used the distribution $\mathcal{N}(0,\sigma^2)$  with different values of $\sigma^2$. The reliability diagrams for the probabilistic forecast of critical region 1 using different load forecast uncertainties are given in Fig. \ref{fig:reliability}. The results show good resolution at the expense of reliability.
\subsubsection{Forecast with probabilistic generation outage}
 We now considered the case of a generating unit outage with a partial loss of capacity, assuming that the maximum capacity of the generator at bus 1 can be reduced to $100$ MW with probability $p$. Other settings were the same as those in the first scenario.

The critical regions under this configuration became $\mathscr{D}^\text{out}=\{(0,100),(100,200)\}$. To predict a future price, we considered all critical regions $\{\mathscr{D}, \mathscr{D}^\text{out}\}$ that load $d_{t+T}$ may fall in, where $\mathscr{D}$ refers to the critical regions in Table \ref{table:3bus_load} under normal conditions. For the outage frequency $p$, we chose two levels: $p=0.01$ and $p=0.1$. The random walk model was adopted to generate load profiles. As benchmarks, both deterministic and certainty equivalence forecasting algorithms also took contingencies into consideration.

\begin{figure} \centering%\vspace{-.5em}
\includegraphics[scale=0.6]{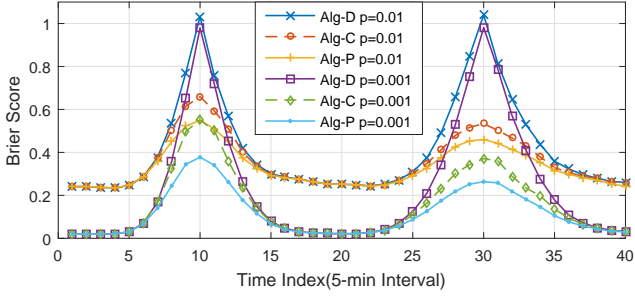}%\vspace{-.25em}
        \caption{Impact of outage frequency $p$. }\label{fig:outage}%\vspace{-1em}
\end{figure}

From Fig. \ref{fig:outage}, the behaviors of all three algorithms with the same outage frequency $p$ were similar to that in the first scenario. Boundary effects and peak asymmetries were observed with contingencies as well. Compare the performances of each algorithm with different outage frequencies: the smaller the outage rate the better the forecast.

\subsection{Case Study: IEEE 118 Bus System}
%\begin{figure}%\hspace{-2em}
%\includegraphics[width=.55\textwidth]{cr118.eps}%\vspace{-.5em}
%\caption{Critical region partition. }\label{fig:cr118}%\vspace{-1em}
%\end{figure}
{
The IEEE 118 bus system was used to show the scalability of the proposed technique and the effectiveness of the heuristic algorithm described in Section \ref{sec:dis}.  The simulations results were focused on the complexity comparison between the proposed techniques and the probabilistic forecast benchmark.

We introduced 12 wind generators (roughly $10\%$ of buses) with $10\sim20\%$ renewable penetration\footnote{By $x\%$ renewable penetration we mean the mean value of total wind generation is $x\%$ of the total electricity load ($4242$ MW in this system). Note that load was assumed to be deterministic in this case.}.  The wind generators were located at bus 25, 26, 90, 91, 100, 103, 104, 105,  107, 110, 111, and 112. The selection of these locations were intended to represent two wind farms, the small one has 2 wind generators, and the large one has 10 wind generators concentrated on a few neighboring buses. All wind generators were assumed to be identical with the maximum capacity of $110$ MW. Denote the wind generation space by the hypercube $\mathscr{W}\in \mathbb{R}^{12}$.  We imposed the maximum capacity of $100$ MW on transmission line 8, 126, and 155. The load profile,  generator capacities and cost functions, and line and bus labels were referred to as in ``case118'' in MATPOWER \cite{zimmerman2011matpower}. Note that the cost function in this system was quadratic, thus the LMP was an affine function of the wind production within each critical region.

\begin{figure}\centering
\includegraphics[scale=0.6]{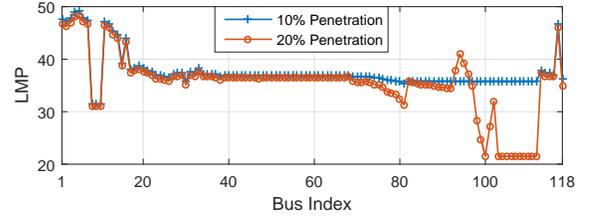}
\caption{LMPs at $10\%$ and $20\%$ renewable penetration levels.}\label{fig:slmp}
\end{figure}

The mean trajectory $\bar{w}(i)$  of each wind generator $i$ was assumed to be linear, for $i=1,\cdots, 12$. In particular, the trajectory starts from $10\%$ penetration level, \textit{i.e.}, $\bar{w}_0(i)=35.35$, at time $0$, and ends at $20\%$ penetration level, \textit{i.e.}, $\bar{w}_{10}(i)=70.70$, at time $10$. The increment was assumed to be constant, \textit{i.e.}, $3.535$ MW. LMP values at $10\%$ and $20\%$ penetration levels are given in Fig. \ref{fig:slmp}. We observed that higher renewable penetration reduced LMP at most buses, but raised the LMPs at bus 93, 94, 95, and 96. The reason of such nonintuitive increase was the congestion of transmission line 155 caused by the increased wind production from the big wind farm. The random walk model was used for the stochastic generation profile. The distribution of the independent noise process $\epsilon_t$ was set to be the standard multivariate Gaussian.
%{\color{blue} Delete this paragraph?} We use this example to describe the forecast process here. Given the mean trajectory $\bar{w}$ of wind production, we used the multiparametric toolbox \cite{MPT3} to partition the hypercube\footnote{The pre-defined hypercube was chosen to cover all possible wind power realizations. In practice, this hypercube can be interpreted as the capacities of each wind generator.} $\mathscr{W} \subseteq\mathbb{R}^{12}$ with range $[0, 110]$ for each dimension. The resulting feasible parameter space was partitioned into $273$ critical regions. For the online forecasting process, given the current realization of stochastic generation $w_t$, we predicted LMP and congestion pattern at time $t+T$. Note that the conditional distribution of $w_{t+T}$ is $\mathcal{N}(\bar{w}_{t+T|t}, T\Sigma)$, where the covariance matrix $\Sigma$ is the identity matrix. The probability of the random generation $w_{t+T}$ falling in each critical region was estimated using the importance sampling technique described in Section \ref{sec:est}: for each critical region $\mathscr{W}_i$, 10,000 samples were drawn from $\mathcal{N}(\bar{v}(\mathscr{W}_i), \Sigma)$, where $\bar{v}(\mathscr{W}_i)$ is the mean of the vertices of $\mathscr{W}_i$, the probability distribution of congestion pattern was obtained from the active/inactive status associated with each transmission line within each critical region.

\begin{figure}\centering
\includegraphics[scale=0.6]{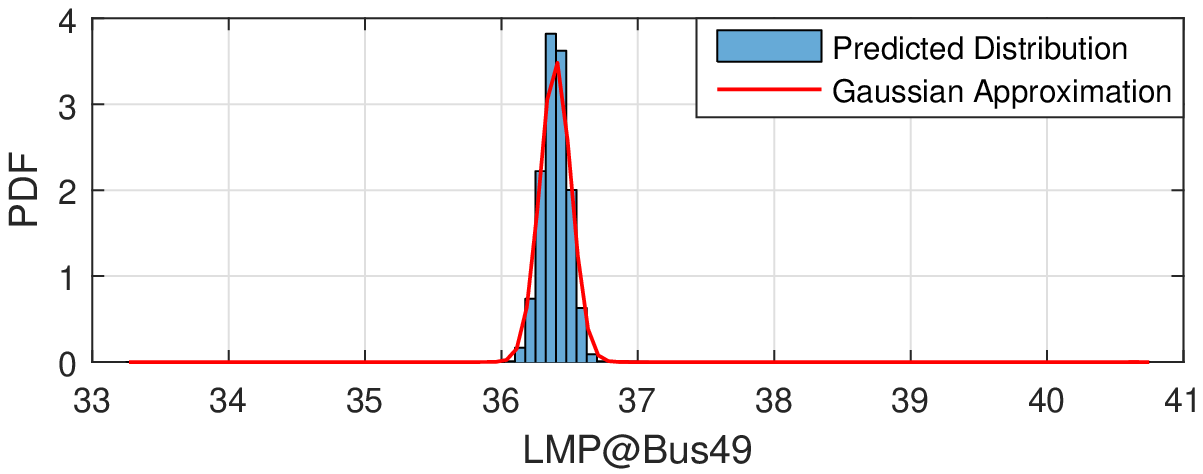}\\
\includegraphics[scale=0.6]{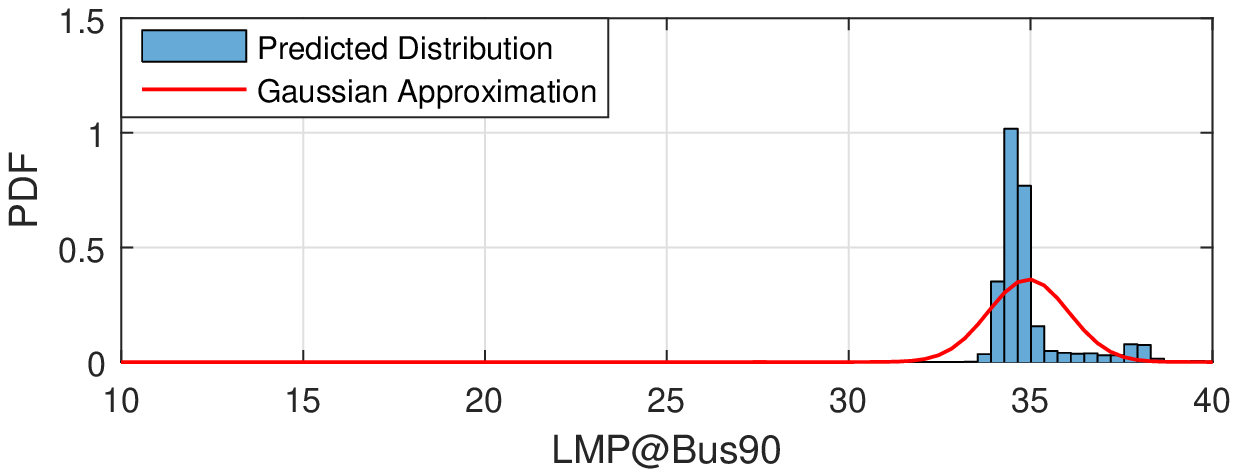}\\
\includegraphics[scale=0.6]{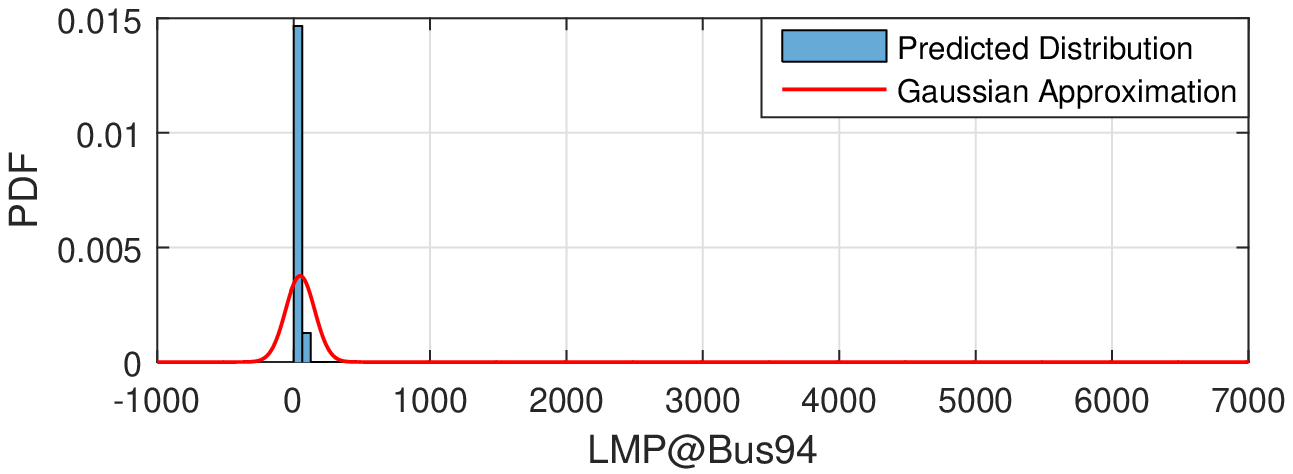}\\
\includegraphics[scale=0.6]{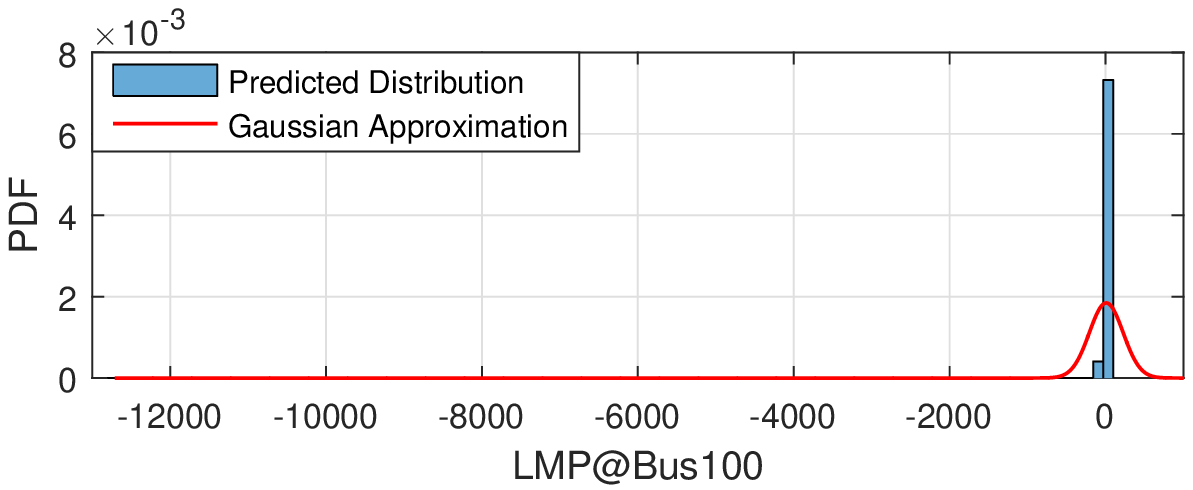}
\caption{Predicted LMP distributions and Gaussian approximations at bus 49, 90, 94, and 100.}\label{fig:lmp_gaussian_approximation}
\end{figure}

Results for the 10-step ahead prediction at time $t=0$ are provided in Fig. \ref{fig:lmp_gaussian_approximation}. The predicted marginal distribution of LMP at bus 49 exhibits the a Gaussian distribution as it is well fitted by the Gaussian approximation with the sample mean and sample variance as distribution parameters. However, such Gaussian characteristics were not observed on the other three buses. According to the distributions of LMP at bus 94 and 100, the extreme values of LMP occasionally appeared as spikes, which were caused by the network congestions.

In the following, we will focus on the complexity performance of the proposed forecasting techniques.  Theoretically, there were at most $2^{115}$  critical regions, because the constraints in the DC-OPF (\ref{dcopf}) for this example included 1 energy balance constraint, 6 transmission constraints, and 108 generator capacity constraints. For the given hypercube $\mathscr{W}$ of the parameter space, there were in total $273$ critical regions. But for the generated 10,000 samples at time $10$, only $17$ critical regions were observed and more than $99\%$ samples fell in $3$ critical regions, as shown in the distribution over the observed $17$ critical regions in Fig. \ref{fig:cr_distribution}.

\begin{figure}\centering
\includegraphics[scale=0.6]{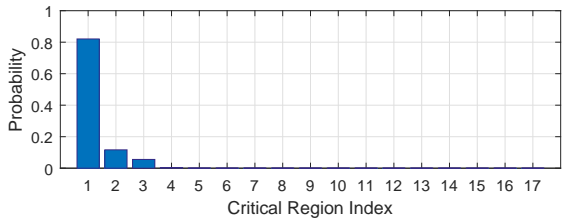}
\caption{Conditional distribution of observed critical regions at time $10$.}\label{fig:cr_distribution}
\end{figure}

Instead of exploring the entire wind production space offline, we implemented the online forecasting algorithm DCRG given in Section \ref{sec:dis}. Fig. \ref{fig:Nopf} shows the comparison of computational cost between the proposed DCRG algorithm, labeled by ``Alg-DCRG'', and the direct Monte Carlo simulations, labeled by `` Alg-MC''. Both algorithms present approximately linear growth in the logarithm scales. But the proposed DCRG algorithm provided more than three orders of magnitude reduction in the number of DC-OPF computations required in the simulation.

\begin{figure}\centering
\includegraphics[scale=0.6]{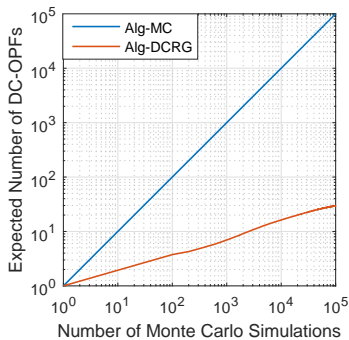}
\caption{The expected number of DC-OPF computations vs. the total number of Monte Carlo simulations.}\label{fig:Nopf}
\end{figure}

Finally, we provide the computation time comparison for the three probabilistic forecasting techniques in Table \ref{table:time}. All computational times were evaluated by implementing the algorithms in Matlab environment with the default ``quadprog'' solver and an external MPT3 \cite{MPT3} toolbox on a desktop with an Intel Core i7-3770 CPU at 3.4 GHz and 8 GB memory. No attempts were made to optimize the efficiency of the algorithms and their simulations. From Table \ref{table:time}, we can conclude that the direct Monte Carlo simulation approach  Alg-MC failed to meet the time constraint for real-time LMP forecasting, since 10,000 samples took more than 14 hours to generate the distribution. The proposed techniques, on the other hand, only took less than half a minute in the online computation, demonstrating the efficiency for the real-time LMP forecasting.

\begin {table}\begin{center}\caption{Computation time (in seconds) for 10,000 samples.}
\begin{tabular}{|c|c|c|c|}  \hline
&Offline&Online & Total\\ \hline
Alg-P&160.60 & 23.32 &   183.92  \\ \hline
Alg-DCRG&---  & 23.59  & 23.59 \\ \hline
Alg-MC&--- & 52022.57  & 52022.57 \\ \hline
\end{tabular}
\label{table:time}\vspace{-2em}
\end{center}\end {table}

}

\section{Conclusion}\label{sec:conclusion}
This paper presents a new methodology for the short-term forecasting of real-time LMP and congestion for system operators. Based on a multiparametric programming formulation of DC-OPF, we have developed an approach that exploits the parametric structure of DC-OPF solutions to obtain conditional distributions of future LMP and congestion.

For system operators, the proposed online forecasting technique provides a new tool for managing operation risks and solving stochastic optimization problems.  See, for example, \cite{ji2015scts}.  For market participants, on the other hand, congestion and LMP forecasts by the operator provide actionable signal for managing flexible resources and demand side management.

{ Currently, there are very few probabilistic forecasting techniques for system operators beside direct Monte Carlo simulations \cite{Min08}.  The approach presented here represents a first step toward online forecasting in large power systems with significant stochastic components. For future work, there is a need to develop computationally tractable techniques, informative performance measure, and a set of practical benchmarks.}

{\appendix[Proof of Theorem \ref{thm:mp_soln_struct}]
\begin{proof} For any parameter $\theta$, denote the optimal primal and dual solutions to MPP (\ref{mplp}) by $x^*(\theta)$ and $y^*(\theta)$ respectively.

 (1) For the MPLP case,  if $\theta$ is in the same critical region as $\theta_0$, they have the same optimal partitions, which means that
\begin{eqnarray}
A_{\mathscr{I}_0}x^*(\theta)-b_{\mathscr{I}_0}-E_{\mathscr{I}_0}\theta&=&\mathbf{0},\label{pfc:active}\\
A_{\mathscr{I}^c_0}x^*(\theta)-b_{\mathscr{I}^c_0}-E_{\mathscr{I}^c_0}\theta&<&\mathbf{0}.\label{pfc:inactive}
\end{eqnarray}
Because MPLP is neither primal nor dual degenerate,  $A_{\mathscr{I}_0}$ has full rank, and
\begin{equation}
x^*(\theta)=A_{\mathscr{I}_0}^{-1}(b_{\mathscr{I}_0}+E_{\mathscr{I}_0}\theta).
\end{equation}
Substituting $x^*(\theta)$ into (\ref{pfc:inactive}), we have $\theta \in \Theta_{\mathscr{I}_0}$.

 (2) For the MPQP case, the first order optimality conditions are given by
 \begin{eqnarray}
 H x^*(\theta)+A^\intercal y^*(\theta)&=&\mathbf{0},\label{kkt:smqp}\\
 Ax^*(\theta)- b-E\theta &\leq& \mathbf{0}, \label{kkt:pf_mqp}\\
 y_i^*(\theta)(A_ix^*(\theta)- b_i-E_i\theta)&=&0, \;\forall i\in\mathscr{J}, \label{kkt:slack_mqp}\\
 y^*(\theta)&\geq& \mathbf{0}.\label{kkt:df_mqp}
 \end{eqnarray}
 From (\ref{kkt:smqp}),
\begin{equation}\label{eqn:x(y)}
x^*(\theta)=-H^{-1}A^\intercal y^*(\theta).
\end{equation}
Substituting the result into (\ref{kkt:slack_mqp}), we have
\begin{eqnarray}
A_iH^{-1}A_i^\intercal y_i^*(\theta)+b_i+E_i\theta&=&0, \;\forall i\in\mathscr{I}_0, \label{eqn:xact}\\
y_i^*(\theta)&=&0,\;\forall i\in \mathscr{I}^c_0.
\end{eqnarray}
By the non-degeneracy assumption, the rows of $A_{\mathscr{I}_0}$ are linearly independent. This implies that $A_{\mathscr{I}_0}H^{-1}A_{\mathscr{I}_0}^\intercal$ is a square full rank matrix. Therefore, from (\ref{eqn:xact}), we solve
\begin{equation}
y_{\mathscr{I}_0}^*(\theta)=-(A_{\mathscr{I}_0}H^{-1}A_{\mathscr{I}_0}^\intercal)^{-1}(b_{\mathscr{I}_0}+E_{\mathscr{I}_0}\theta)
\end{equation}
and substitute $y_{\mathscr{I}_0}^*(\theta)$ and $y_{\mathscr{I}_0^c}(\theta)$ into (\ref{eqn:x(y)}) to obtain
\begin{equation}\label{eqn:x(theta)_mqp}
x^*(\theta)=H^{-1}A_{\mathscr{I}_0}^\intercal (A_{\mathscr{I}_0}H^{-1}A_{\mathscr{I}_0}^\intercal)^{-1}(b_{\mathscr{I}_0}+E_{\mathscr{I}_0}\theta).
\end{equation}

Substituting $x^*(\theta)$ from (\ref{eqn:x(theta)_mqp}) in the primal feasibility conditions (\ref{kkt:pf_mqp}) gives $\mathscr{P}_p$ and substituting $x^*(\theta)$ from (\ref{eqn:x(theta)_mqp}) in the dual feasibility condition (\ref{kkt:df_mqp}) gives $\mathscr{P}_d$. We therefore have $\theta \in \Theta_{\mathscr{I}_0}$.
\end{proof}
}
\section*{Acknowledgement}
The authors thank Kory Hedman for pointing out the need for incorporating time varying constraints and anonymous reviewers for their detailed comments.

\bibliographystyle{IEEEtran}
{\bibliography{reference}}
\end{document}